\documentclass[final]{aipproc}

\layoutstyle{6x9}

\usepackage{wasysym,amssymb}

\newcommand{\be}{\begin{equation}}
\newcommand{\ee}{\end{equation}}
\newcommand{\bea}{\begin{eqnarray}}
\newcommand{\eea}{\end{eqnarray}}
\newcommand{\eref}[1]{(\ref{#1})}

\newcommand{\D}{\mathcal{D}}
\newcommand{\Or}{\mathcal{O}}

\newcommand{\m}{\, \mathrm{m}}
\newcommand{\cm}{\, \mathrm{cm}}
\newcommand{\s}{\, \mathrm{s}}
\newcommand{\km}{\, \mathrm{km}}
\newcommand{\kg}{\, \mathrm{kg}}
\newcommand{\g}{\, \mathrm{g}}

\begin{document}

\title{Solar System Constraints on \\ Gauss-Bonnet Dark Energy}


\author{ Stephen C. Davis}{
  address={Lorentz Institute, Postbus 9506, 2300 RA Leiden, 
The Netherlands}
}

\begin{abstract}
Quadratic curvature Gauss-Bonnet gravity may be the solution to the
dark energy problem, but a large coupling strength is required. This
can lead to conflict with laboratory and
planetary tests of Newton's law, as well as light bending. The
corresponding constraints are derived. If applied directly to
cosmological scales, the resulting bound on the density fraction is
$|\Omega_{GB}| \lesssim 3.6 \times 10^{-32}$. 
\end{abstract}

\maketitle

\section{Gauss-Bonnet gravity and the solar system}

Corrections to Einstein gravity, such as the
string-motivated Gauss-Bonnet term 
$\mathcal{L}_{GB} = R^2 - 4R_{\mu\nu} R^{\mu\nu} + R_{\mu\nu\rho\sigma}
R^{\mu\nu\rho\sigma}$ could explain the current accelerated expansion
of our universe. On its own, in four dimensions, the Gauss-Bonnet
term does not contribute to the gravitational field
equations. Coupling it to a scalar field will produce a non-trivial
effect, which could act as effective dark energy. Including the
corresponding higher order scalar kinetic terms, we obtain the
 ghost-free, quadratic curvature, gravitational Lagrangian
\be
\mathcal{L} = \sqrt{-g} \, \biggl\{R - (\nabla \phi)^2 
+ \xi_1 \mathcal{L}_{GB}
+ \xi_2 G^{\mu\nu}\nabla_{\! \mu} \phi \nabla_{\! \nu} \phi
+\xi_3 (\nabla \phi)^2\nabla ^2\phi +\xi_4  (\nabla \phi)^4
 \biggr\}  \, .
\ee

The gravity modifications will not only be felt at cosmological
scales, but also within the solar system where high precision
gravitational measurements have been performed. The fields there are
relatively weak and slowly varying, allowing us to use the approximate
post-Newtonian metric~\cite{will}
\be
ds^2= -(1+2 \Phi/c^2)(c\, dt)^2  + (1-2 \Psi/c^2) dx^i dx_i 
+ \Or(\epsilon^{3/2}) \, ,
\ee
with $\Phi, \Psi \sim c^2 \epsilon$, and $\partial_t \sim \epsilon^{1/2}$. 
We take $\phi =\phi_0 + \Or(\epsilon)$, with $\phi_0$ a constant.
For standard Einstein gravity $\Phi = \Psi = - G m/r$. We find the
expansion parameter satisfies $\epsilon \lesssim 10^{-5}$. 

A perturbative analysis of the gravitational field equations can now
be performed. However, it should be noted that if the field-dependent couplings
$\xi_i(\phi)$ are to produce cosmological acceleration, they must be
large. With this in mind we make no assumptions on the relative
magnitude of $\xi_i(\phi_0)$ and $\epsilon$. 
For simplicity we will assume the $\xi_i(\phi)$, and
all of their derivatives, are of the same order. This is the case, for
example, when $\phi$ arises from a toroidal compactification of a higher
dimensional space~\cite{us}. 

To leading order in $\epsilon$, the scalar field equation is
\be
c^4 \Delta \phi=  - 4\xi_1' \D(\Phi,\Psi) 
+ \Or(\epsilon^2,\xi_i \epsilon^3/r^2) \, ,
\ee
where $\xi_1' = \partial \xi_1/\partial \phi$, evaluated at $\phi=\phi_0$. The 
Einstein equations take the form
\bea
 && \Delta \Phi = 4\pi G \rho_m 
- 2\xi_1'  \D(\Phi+\Psi,\phi) 
+ \Or(\epsilon^2,\xi_i \epsilon^3/r^2) \, ,
\\ &&
 \Delta \Psi = 4\pi G \rho_m 
- 2\xi_1'  \D(\Psi,\phi) 
+ \Or(\epsilon^2,\xi_i \epsilon^3/r^2) \, ,
\eea
with $\rho_m$ the matter energy density in
the solar system. We have defined  the operators
\be
\Delta X = \sum_i X_{,ii} \  , \qquad
\D(X,Y) = \sum_{i,j} X_{,ij} Y_{,ij} - \Delta X \Delta Y \, .
\ee
with $i,j=1,2,3$. To leading order, the Gauss-Bonnet term is 
$\mathcal{L}_{GB} = 8 \D(\Phi,\Psi)$.

To agree with observation, the solution to the above equations must be
close to the usual Einstein gravity results. We can therefore assume $\Phi = -G
m/r + \Or(\xi_i)$, etc., from which we obtain, to leading order~\cite{full}
\be
\phi = \phi_0 - 2\xi_1' \frac{(G m)^2}{c^4 r^4}
\, , \quad
\Phi = -\frac{G m}{r}
- \frac{64}{7} \frac{\xi_1'^2 (G m)^3}{c^4 r^7}
\, , \quad
\Psi = -\frac{G m}{r}
-\frac{32}{7} \frac{\xi_1'^2 (G m)^3}{c^4 r^7}
\, .
\label{dPhi}
\ee
We see there are  mass-dependent, $1/r^{7}$ corrections, which are not covered
by the usual parametrised post-Newtonian formalism~\cite{will}.
This is in agreement with~\cite{gilles}, but not~\cite{GBPPN} (which
does not allow for the possibility that the couplings $\xi_i$, could be large).

\subsection{Planetary motion}

Planets in our solar system
experience a gravitational acceleration $g_{\rm acc} = -G m/r^2$,
resulting in elliptical orbits
with period  $2 \pi \sqrt{a^3/(G m)}$, where $a$ is the semi-major axis of the
planet and $m$ is the sun's mass. Corrections to the Newtonian potential
alter the effective mass felt by the
planets~\cite{anderson,SSlambda}. From~\eref{dPhi} we obtain~\cite{full}
\be
g_{\rm acc}(r) = -  \frac{d\Phi}{dr}
=-\frac{G m}{r^{2}}\left[1
-\frac{64\xi_1'^2 r_g^2}{r^{6}}\right]
\equiv -\frac{G (m + \delta m)}{r^{2}}
\ee
where $r_{g} \equiv G m/c^2 \approx 1.5\km$ is gravitational radius of the
sun. To agree with observation, the correction must be
smaller than the uncertainty in $a$, so $\delta m/m < 3
\delta a/a$.

The strongest bound
comes from 
Mercury (with $a\approx 5.8\times 10^{7}\km$ and  $\delta a \approx 0.11
\m$~\cite{pitjeva})
\be
|\xi'_1|\lesssim 
\left. \frac{\sqrt{3a^{5} \delta a}}{8r_{g}} \right|_{\mercury}
\approx 3.8\times 10^{16}\km^{2} \, .
\label{GBcon1}
\ee

Applying this directly to Gauss-Bonnet density fraction~\cite{us}
in cosmology, we find
\be
|\Omega_{GB}| = \left|4 \xi'_1 H \frac{d\phi}{dt}\right| \lesssim 8.8 \times
10^{-30} 
\ee
if $d \phi/dt \sim H$, and if $\xi'_1(\phi)$ has comparable
values on local and cosmological scales. This value is far short the
$0.7$ required to solve the dark energy problem.

For a cosmological constant $\Phi = -G m/r -r^2 c^2 \Lambda/6+\cdots$. 
The corresponding bound  comes
from Mars~\cite{SSlambda} ($a \approx 2.3 \times 10^8 \km$,
 $\delta a \approx 0.66 \m $~\cite{pitjeva}) and is
\be
|\Lambda| \lesssim \left. \frac{9 r_g \delta a}{a^4}\right|_{\mars} 
\approx 1.2\times 10^{-34}\km^{-2} \, .
\ee
This implies $\Omega_\Lambda = \Lambda/(3H^2) \lesssim 7.3\times 10^{11}$, 
which is vastly weaker than the corresponding cosmological
constraint ($\Omega_\Lambda \lesssim 1$).

\subsection{Cassini spacecraft}

An even stronger constraint is obtained from signals between the earth
(at $r_\oplus \approx 1.5 \times 10^{8}\km$) and 
Cassini spacecraft (at $r_{e} \approx 1.3 \times 10^{9}\km$) as it
travelled to Saturn. For a round trip, the sun's gravitational field
produces a time delay in the signals of~\cite{full}
\be
c \Delta t= 2\int_{\rm ray}
\left(\sqrt{\frac{g_{xx}}{-g_{tt}}}-1\right) dx 
\approx -2\int_{\rm ray}
(\Phi+\Psi) \, dx 
\approx 4 r_g \ln \frac{r_\oplus r_e}{4b^2}
+\frac{1024 \xi_1'^2 r_g^3}{b^6} \, ,
\label{Dt}
\ee
where the impact parameter $b$, is the smallest value of $r$ on the
signal's path. In 2002 it fell to its lowest value, 
$b \approx 1.1 \times10^{6}\km$. 

Rather than directly measure $\Delta t$, the Cassini experiment
actually found the frequency shift in the signal~\cite{cassini}
\be
y_{\rm gr} = \frac{d \Delta t}{dt} 
\approx \frac{d \Delta t}{db} \frac{db}{dt} 
 = -\frac{10^{-5} \s}{b}\frac{db}{dt}
\left[2 +(2.1\pm2.3) \times 10^{-5} \right] \, .
\label{ygr}
\ee
Requiring the Gauss-Bonnet correction~\eref{Dt} to be within the
measured range~\eref{ygr} implies the bounds
\be
|\xi'_1|\lesssim 1.6 \times 10^{14}\km^{2} \, ,
\qquad |\Omega_{GB}| \lesssim 3.6 \times 10^{-32}\, .
\ee

\subsection{A table-top laboratory test of Newton's law}

Laboratory tests will also constrain modified gravity, as we will
illustrate with the experiment described in~\cite{hoskins}. It consists
of a $60\cm$ copper bar, suspended at its midpoint by a tungsten wire. Two 
$7.3 \kg$ masses $105 \cm$ from the bar produce a torque $N_{105}$ on
the bar, and an $m \approx 43 \g$ mass $5 \cm$ to the side of bar produces a
comparable torque $-N_5$. By changing the positions of the
masses, the ratio $R = N_{105}/N_5$ was determined and compared to theory
\be
\delta_R = \frac{R_{\rm expt}}{R_{\rm Newton}}-1 = (1.2 \pm 7) \times 10^{-4}
\ .
\ee

The Gauss-Bonnet term affects all the masses, and gives cross-terms due its
non-linearity. However we can ignore these complications, and just
use~\eref{dPhi} for the mass $m$, since it gives the dominant
correction.  A mass at $\vec X=(X,Y,Z)$ produces a torque
\be
N = \int_{\rm bar} d^3x \, (\vec x \wedge \vec F)_z
\propto \int_{\rm bar} d^3x \,
\frac{y X -x Y}{r} \left. 
\frac{d\Phi}{dr}\right|_{r  = |\vec X - \vec x|} \, .
\ee
We find $\delta N_5/N_5 \approx - 0.003 (G m \xi_1')^2 c^{-4}
\cm^{-6}$. 
Requiring $\delta N_5/N_5 < \delta_R$, gives the bound $|\xi_1'|
\lesssim  1.3 \times 10^{16} \km^2$, which is comparable to the planetary
constraint~\eref{GBcon1}.

\section{Discussion}

Extrapolating solar system constraints to cosmological scales
suggests that the density fraction $\Omega_{GB}$ is far too small to
solve the dark energy problem. However our analysis features many
assumptions, which while credible, could be violated and thus offer a
way round the constraints. Clearly at least one of them must be broken
if Einstein-Gauss-Bonnet gravity is to explain the acceleration of our
universe.

In particular, we applied solar system results directly to
cosmological scales. This assumes no significant spatial or temporal
evolution of the field $\phi$. Significant variation in the couplings
$\xi_i$ seems to offer the best way to save Gauss-Bonnet dark
energy. Another possibility is that $\phi$ couples
differently to dark matter and baryons, which will also break the relation
between the two scales.

Instead, it may be that our assumptions on the form of the
theory should be changed. The scalar field could be coupled directly
to the Einstein-Hilbert term, as in Brans-Dicke gravity. Additionally,
the couplings $\xi_i$ and their derivatives could be of different
orders. Both these changes open up the possibility of the corrections to
Einstein gravity cancelling within the solar system. Alternatively $\phi$
could be given a large mass, which would suppress the quadratic
curvature effects, as they operate via the scalar field. However this
is also likely to inhibit acceleration.

\begin{theacknowledgments}
I am grateful to the  Netherlands Organisation for Scientific
Research (NWO) for financial support, to my collaborators
C. Charmousis and L. Amendola, and of course to the conference
organisers themselves. 
\end{theacknowledgments}

\end{document}